\documentclass[10pt,preprint]{aastex}

\shorttitle{Solar Microwave Bursts with Fine Structures}
\shortauthors{Tan et al.}

\begin{document}

\title{Microwave Zebra Pattern Structures in the X2.2 Solar Flare on Feb 15, 2011}
\author{Baolin Tan\altaffilmark{1}, Yihua Yan\altaffilmark{1}, Chengming Tan\altaffilmark{1}, Robert Sych\altaffilmark{1,2}, Guannan Gao\altaffilmark{3,1}}
\affil{$^{1}$Key Laboratory of Solar Activity, National
Astronomical Observatories \\ Chinese Academy of Sciences, Beijing
100012, China}

\affil{$^{2}$ Institute of Solar-Terrestrial Physics of Siberian Branch of Russian Academy of Sciences\\
126a Lermontov Street, Irkutsk, 664033, Russia}

\affil{$^{3}$ National Astronomical Observatories/Yunnan
Astronomical Observatories \\ Chinese Academy of Sciences, Kunming
650011, Yunnan Province, China}

\altaffiltext{1}{Datun Road A20, Chaoyang District, Beijing,
100012, China.}

\begin{abstract}

Zebra pattern structure (ZP) is the most intriguing fine structure
on the dynamic spectrograph of solar microwave burst. On 15
February 2011, there erupts an X2.2 flare event on the solar disk,
it is the first X-class flare since the solar Schwabe cycle 24. It
is interesting that there are several microwave ZPs observed by
the Chinese Solar Broadband Radiospectrometer (SBRS/Huairou) at
frequency of 6.40 $\sim$ 7.00 GHz (ZP1), 2.60 $\sim$ 2.75 GHz
(ZP2), and the Yunnan Solar Broadband Radio Spectrometer
(SBRS/Yunnan) at frequency of 1.04 -- 1.13 GHz (ZP3). The most
important phenomena is the unusual high-frequency ZP structure
(ZP1, up to 7.00 GHz) occurred in the early rising phase of the
flare, and there are two ZP structure (ZP2, ZP3) with relative low
frequencies occurred in the decay phase of the flare. By
scrutinizing the current prevalent theoretical models of ZP
structure generations, and comparing their estimated magnetic
field strengths in the corresponding source regions, we suggest
that the double plasma resonance model should be the most possible
one for explaining the formation of microwave ZPs, which may
derive the magnetic field strengths as about 230 -- 345 G, 126 --
147 G, and 23 -- 26 G in the source regions of ZP1, ZP2, and ZP3,
respectively.

\end{abstract}


\section{Introduction}

After a super-long quietness, the Sun begins a new Schwabe cycle.
On 15 February 2011, there erupts an X2.2 flare event in active
region NOAA AR11158 on the solar disk, which was the first X-class
GOES flare of the current solar Schwabe cycle 24. Several
instruments observed this event, including Solar Dynamics
Observatory (SDO), Nobeyama Radio Polarimeter (NoRP) and Nobeyama
Radio Heliograph (NoRH), RHESSI, Chinese Solar Broadband Radio
Spectrometer (SBRS/Huairou), and Yunnan Solar Broadband Radio
Spectrometer (SBRS/Yunnan). Especially from the broadband radio
spectrogram observations at SBRS/Huairou and SBRS/Yunnan, strong
microwave bursts with spectral fine structures, such as zebra
patterns (ZP) and some other fine structures are registered. The
most interesting and important phenomena is the microwave ZP
structures. We have registered three ZP structures at different
frequency bands and in different phases of the flare. We will
focus on investigating these ZP structures in this work.

ZP is a fine spectral structure superposed on the solar radio
broadband type IV continuum spectrogram, which consists of several
almost parallel and equidistant stripes. Most often, ZP structures
are observed in meter and decimeter frequency range(Slottje, 1972,
etc). In microwave range such structure is very rare. From the
publications heretofore, the highest frequency at which ZP
structure has been observed is about 5.70 GHz (Altyntsev et al,
2005). Recently, Chernov et al (2011) find out some ZP evidences
at frequency of 5.70 $\sim$ 7.20 GHz from the observations of a
solar flare on May 29, 2003 obtained at SBRS/Huairou. However, it
is far infrequent to observe ZP structure at such high frequency
range. Additionally, ZP structures are typically occurred around
the impulsive phase and/or decay phase of solar flares. It is very
rare to find ZP structure in the early rising phase of solar
flare.

The formation mechanism of ZP structures has been a subject of
widely discussion for more than 40 years. The historical
development of observations and theoretical models are assembled
in the review of Chernov (2006), Zlotnik (2009), and so on. There
are several theoretical models to interpret ZP structures, which
are mainly developed to apply to meter and decimeter wavelengths
(Rosenberg, 1972; Kuijpers, 1975; Zheleznyakov \& Zlotnik, 1975;
Chernov, 1976, 1990; LaBelle, et al, 2003; Kuznetsev, 2005;
Ledenev, et al, 2006; Tan, 2010; etc). These models can be
classified simply into three classes:

(1) Isogenous models, which proposed that all the stripes in a ZP
structure are generated from a small compact source.

The model of Bernstein mode (BM model) is the first one to
interpret the formation of ZP structure. The emission mechanism is
the nonlinear coupling between two Bernstein modes, or Bernstein
mode and other electrostatic upper hybrid waves. The electrons
with non-equilibrium distribution over velocities perpendicular to
the magnetic field are located in a small source, where the plasma
is weakly magnetized ($f_{pe}\gg f_{ce}$), and the magnetic field
is uniform. These electrons excite longitudinal electrostatic
waves at frequency of the sum of so-called Bernstein modes
frequency $sf_{ce}$ and the upper hybrid frequency $f_{uh}$. The
BM excitation occurs in relatively narrow frequency band. The
emission frequency (Rosenberg, 1972; Chiuderi et al, 1973; Zaitsev
and Stepanov, 1983) is:

\begin{equation}
f=f_{uh}+sf_{ce}\approx f_{pe}+sf_{ce}.
\end{equation}

Here, $f_{pe}$ is the electron plasma frequency, $f_{ce}$ the
electron gyro-frequency, $s$ is harmonics number. This model
presents the frequency separation between the adjacent zebra
stripes just as the electron gyro-frequency: $\Delta f=f_{ce}$
(Zheleznyakov and Zlotnik, 1975).

(2) Heterogenous models, which proposed that zebra stripes in a
structure are generated from some extended source regions in the
magnetic flux tube (Kuijpers, 1975; Fomichev and Fainshtein, 1981;
Mollwo, 1983; Ledenev, Yan, and Fu, 2001; Chernov et al, 2005;
Altyntsev et al, 2005).

One of the important heterogenous model is based on the plasma
waves interact with whistler waves (Chernov, 1996, 2006), called
as whistler wave model (WW model). The coupling of plasma wave and
whistler wave can operate in different conditions: when whistlers
generate at the normal Doppler cyclotron resonance they can escape
along the magnetic loop and yield fiber bursts; when whistlers
generate at the anomalous Doppler cyclotron resonance under large
angles to the magnetic field they may form standing wave packets
in front of the shock wave, and when the group velocity of
whistlers is approximated to the shock velocity, a ZP structure
with slow oscillating frequency drift will appear. The whistler
wave group velocity peaks at whistler frequency $f_{w}\sim
0.25f_{ce}$. The frequency separation $\Delta f$ between adjacent
zebra stripes is about 2 times of whistler frequency: $\Delta
f\sim 2f_{w}$, and then we may obtain: $f_{ce}\sim2\Delta f$.

The most developed heterogenous model for ZP structure generation
is called double plasma resonance model (DPR model), which explain
ZP structure in a natural way (Pearlstein, et al, 1966;
Zheleznyakov \& Zlotnik, 1975; Berney \& Benz, 1978; Winglee \&
Dulk, 1986; Zlotnik et al, 2003; Yasnov \& Karlicky, 2004;
Kuznetsov \& Tsap, 2007). This model proposed that enhanced
excitation of plasma waves occurs at some resonance levels where
the upper hybrid frequency coincides with the harmonics of
electron gyro-frequency in the inhomogeneous flux tube:

\begin{equation}
f_{uh}=(f_{pe}^{2}+f_{ce}^{2})^{1/2}=sf_{ce}
\end{equation}

The emission frequency is dominated not only by the electron
gyro-frequency, but also by plasma frequency. When the emission
generates from the coalescence of two excited plasma waves, the
polarization may be very weak, the emission frequency is $f\approx
2f_{pe}\approx 2sf_{ce}$, and the frequency separation between the
adjacent zebra stripes is $\Delta
f=\frac{2sf_{ce}H_{b}}{|sH_{b}-(s+1)H_{p}|}$. Here,
$H_{b}=f_{B}(df_{B}/dr)^{-1}=B(dB/dr)^{-1}$ and
$H_{p}=f_{pe}(df_{pe}/dr)^{-1}=2n_{e}(dn_{e}/dr)^{-1}=2H_{n}$.
$H_{b}$ and $H_{n}$ are the scale heights of magnetic field $B$
and the plasma density $n_{e}$ in the source regions,
respectively. For $f_{ce}\ll f_{pe}$ and $s\gg 1$, we may get:

\begin{equation}
\Delta f\approx\frac{2H_{b}}{|H_{b}-2H_{n}|}f_{ce}
\end{equation}

When the emission generates from coalescence of an excited plasma
wave and a low frequency electrostatic wave, the polarization will
be strong, the emission frequency is $f\approx f_{pe}\approx
sf_{ce}$, and the frequency separation between the adjacent zebra
stripes is $\Delta f=\frac{sf_{ce}H_{b}}{|sH_{b}-(s+1)H_{p}|}$.
For $f_{ce}\ll f_{pe}$ and $s\gg 1$, we may get:

\begin{equation}
\Delta f\approx\frac{H_{b}}{|H_{b}-2H_{n}|}f_{ce}
\end{equation}

Equ. (3) and (4) indicate that the frequency separation between
the adjacent zebra stripes is dominated by both the scale heights
of the magnetic field and plasma density.

Ledenev et al (2001) proposed another heterogenous model to
interpret the formation of ZP structures. This model (named as
Ledenev model) suggests that the emission generated from an
anisotropic energetic electron beam at low cyclotron harmonics in
a significantly inhomogeneous magnetic field. The electron beam is
formed as a result of fast local energy release in corona, such as
magnetic reconnection, it will excite longitudinal waves at the
normal Doppler resonance. The coalescence of upper-hybrid waves
and low-frequency longitudinal waves ($U+L\rightarrow T$) can
produce electromagnetic waves ($T$). This model gives the
cyclotron frequency harmonics ratio as (Sawant et al, 2002):

\begin{equation}
\frac{f_{s}}{f_{s+1}}=\frac{f_{pe,s}}{f_{pe,s+1}}[\frac{s^{3}(s+2)}{(s+1)^{3}(s-1)}]^{1/2}
\end{equation}

$f_{pe,s}$ and $f_{pe,s+1}$ are the electron plasma frequencies
corresponding to the level of $s$ and $s+1$ harmonics. If density
changes slowly in source regions, then $f_{pe,s}\approx
f_{pe,s+1}$, and the frequency ratio of the cyclotron frequency
harmonics can be presented roughly as: $s=2$, $f_{2}/f_{3}\approx
1.089$; $s=3$, $f_{3}/f_{4}\approx 1.027$; $s=4$,
$f_{4}/f_{5}\approx 1.012$; $s=5$, $f_{5}/f_{6}\approx 1.006$,
etc. In this regime, the magnetic field strength in source region
can be estimated: $B\approx\frac{f_{max}}{2.8\times10^{6}s}, (G)$,
here $f_{max}$ is the frequency on the zebra stripe with maximum
harmonic number $s$.

(3) Propagating model, which proposed that ZP stripes are formed
in the propagating processes after emitted from its source region.
The interference model which suggests that ZP is possibly formed
from some interference mechanism in the propagating processes
(Ledenev, Yan, and Fu, 2006). They suppose that there may exist
some inhomogeneous layers with small size in solar coronal plasma,
and such structure will change the radio wave into direct and
reflected rays. When the direct and reflected rays meet at some
places, interference will take place and form ZP structure.
However, this model needs a structure with great number of
discrete narrow-band sources in small size. Tan (2010) proposed
that such structure may exist in the current-carrying flaring
plasma loop, where the tearing-mode instability forms a great
number of magnetic islands which may provide the main conditions
for the interference mechanism.

ZP is the most intriguing fine structure on the dynamic
spectrogram of solar radio observations, especially the microwave
ZP structures, which may provide the original information of the
solar flaring region, such as the magnetic field, particle
acceleration, and the plasma parameters, etc. In this work, the
observations and data analysis of the microwave ZP structures
associated with the X2.2 flare are presented in Section 2, Section
3 is the physical discussion on the microwave ZP structures,
especially the estimations of the magnetic field strengths from
different ZP models, and pinpoint the reasonable interpretation of
the ZP structures. At the final, some conclusions are summarized
in Section 4.

\section{Observations and Data Analysis}

The active region AR11158 appeared near the center of the solar
disk in the rising phase of the current solar cycle 24. It
developed from a simple $\beta$- to a complex
$\beta\gamma\delta$-configuration rapidly during 12 --21 February
2011. It produced an X2.2 flare on 15 February 2011. This flare
was a two-ribbon white-light flare showed in the image of HMI/SDO,
and accompanied with a large coronal mass ejection which launched
just towards the Earth. From the soft X-ray emission obtained by
GOES, the flare starts at 01:46 UT, reaches to the maximum at
01:56 UT, and ends at 02:07 UT (Maurya, Reddy, \& Ambastha, 2011).

In this work, we focused on the observations obtained from
SBRS/Huairou and SBRS/Yunnan. SBRS/Huairou is an advanced solar
radio telescope with super high cadence, broad frequency
bandwidth, and high frequency resolution, which can distinguish
the super fine structures from the spectrogram (Fu et al 1995; Fu
et al 2004; Yan et al, 2002). It includes 3 parts: 1.10 $\sim$
2.06 GHz (with the antenna diameter of 7.0 m, cadence of 5 ms,
frequency resolution of 4 MHz), 2.60 $\sim$ 3.80 GHz (with the
antenna diameter of 3.2 m, cadence of 8 ms, frequency resolution
of 10 MHz), and 5.20 $\sim$ 7.60 GHz (share the same antenna of
the second part, cadence of 5 ms, frequency resolution of 20 MHz).
The antenna points to the solar center of automatically controlled
by a computer. The spectrometer can receive the total flux density
of solar radio emission with dual circular polarization (left- and
right handed circular polarization), and the dynamic range is 10
dB above quiet solar background emission. And the observation
sensitivity is: $S/S_{\bigodot}\leq 2\%$, here $S_{\bigodot}$ is
quiet solar background emission. Similar to other several
spectrometers, such as Phoenix (100 $\sim$ 4000 MHz, Benz et al,
1991), Ond\'rejov (800 $\sim$ 4500 MHz, Jiricka et al, 1993) and
BBS (200 $\sim$ 2500 MHz, Sawant et al, 2001), SBRS/Huairou have
no spatial resolution. However, as the Sun is a strong radio
emission source, a great deal of works (e.g. Dulk, 1985, etc) show
that the microwave bursts received by spectrometers are always
coming from the solar active region when the antenna points to the
Sun. Additionally, we also adopt the observations at SBRS/Yunnan,
which operating frequency band is in 0.65 $\sim$ 1.50 GHz, with a
spectral resolution of 1.4 MHz, and time resolution of 80 ms, by
using a 10-meter diameter antenna.

During the X2.2 flare, SBRS/Huairou has two parts (2.60 $\sim$
3.80 GHz and 5.20 $\sim$ 7.60 GHz) on duty, SBRS/Yunnan and NoRP
also obtained perfect radio observations. Fig.1 presents the
profiles of the microwave flux at frequencies of 1.00, 2.00, 2.85,
3.75, 6.82, 9.40, 17.00, and 35.00 GHz in 01:45 $\sim$ 02:30 UT
observed at SBRS/Huairou and NoRP. As a comparison, the profiles
of GOES soft X-ray intensities at wavelengths of 1 $\sim$ 8 \AA
(GOES8) and 0.5 $\sim$ 4 \AA (GOES4) are plotted in panel (9) of
Fig.1. Additionally, the plasma temperature derived from the ratio
of GOES soft X-ray emission fluxes at the two wavelength bands
(Thomas, Starr, \& Crannell, 1985) is also over-plotted. Here, we
find that the maximum temperature (at about 01:53 UT) is prior to
the GOES flare peak (01:56 UT) for about 3 minutes.

From the scrutinizing of the microwave spectrogram, we find that
there is a strong microwave type IV burst with spectral continuum
in the frequency range of 0.65 -- 1.50, 2.60 -- 3.80, 5.20 -- 7.60
GHz, and the corresponding microwave enhancements are also
occurred at frequency of 9.40, 17.00, 35.00 GHz. Superposed on
these continuum enhancements, we identify many kinds of fine
structures, such as type III bursts, spike bursts, patches, fast
quasi-periodic pulsations (QPP), fibers and ZP structures. The
main part of the microwave bursts at low frequencies occurred
after the GOES flare peak. However, at the higher frequency range,
the main part of the microwave bursts are prior to that of lower
frequencies. Most of the fine structures are occurred at the
moderate frequency band of 2.60 -- 3.80 GHz after the GOES flare
peak, only one segment of microwave ZP structure occurred prior to
the GOES flare peak (marked as ZP1 in Fig.1) which frequency is
around 6.70 GHz.

\begin{figure}
\begin{center}
 \includegraphics[width=15 cm]{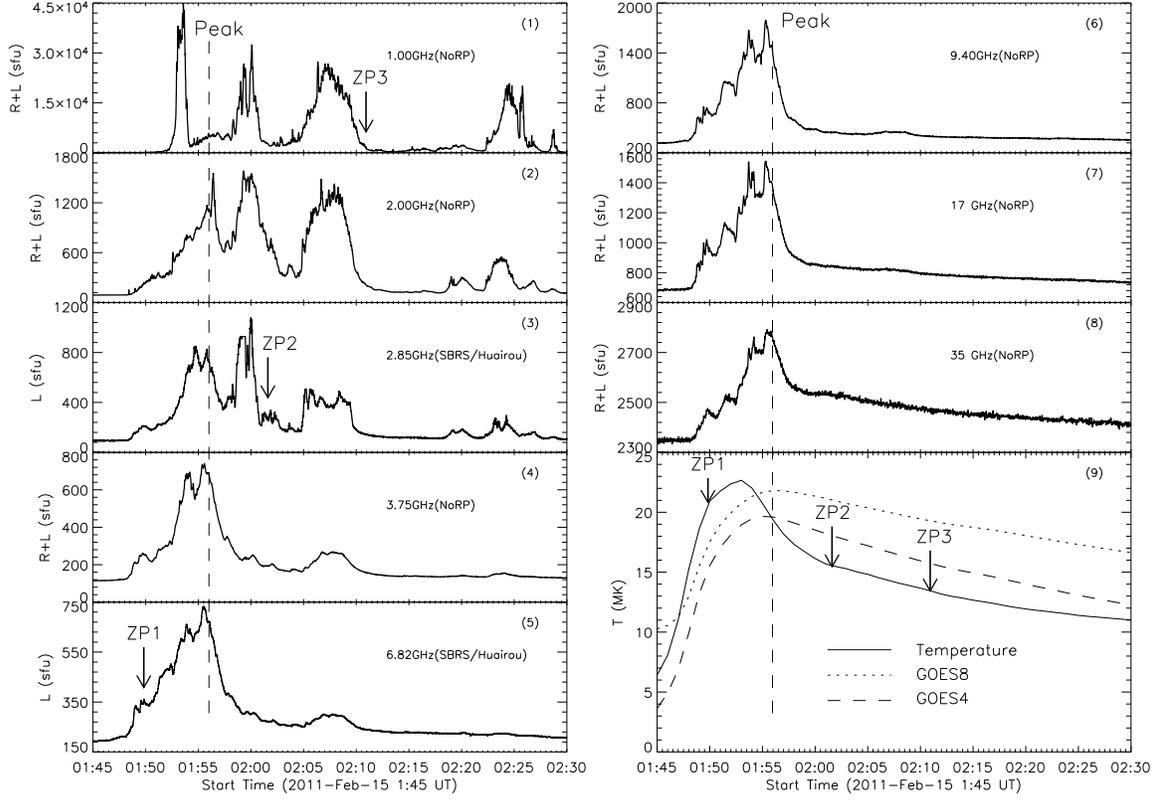}
  \caption{Panel (1) -- (8) are profiles of the microwave flux at frequencies of 1.00, 2.00, 2.85, 3.75, 6.82, 9.40, 17, and 35 GHz in
  01:45 $\sim$ 02:30 UT, 15 Feb. 2011. NoRP indicates the observation of Nobeyama Radio Polarimeter, and SBRS/Huairou indicates the observation of the
  Chinese Solar Broadband Radio Spectrometer (SBRS/Huairou). Panel (9) shows the GOES soft X-ray profiles at 1 $\sim$ 8 \AA (GOES8) and 0.5 $\sim$ 4 \AA (GOES4), and the
  plasma temperature induced from the GOES soft X-ray emission. The arrows indicate the positions of ZP structures on the profiles. }
\end{center}
\end{figure}

In this work, our interesting is mainly focused on microwave ZP
structures. There are 3 segments of microwave ZP structures
registered on the microwave spectrogram: the first one occurred at
frequency of about 6.70 GHz (ZP1), the second one is at about 2.68
GHz (ZP2), and the third one is at about 1.08 GHz (ZP3). With a
simple glance, we find that the higher frequency of ZP structure
appears at the earlier flare phase, while the lower frequency of
ZP structures take place in the later flare phase. The following
is the details.

(1) Microwave ZP Structure at 6.40 $\sim$ 7.00 GHz

In the frequency of 5.20 $\sim$ 7.60 GHz, the microwave emission
always behaves as continuous type IV burst, and the fine structure
is very rare during solar flares. However, the most interesting
phenomenon is an ZP structure occurred at frequency of around 6.70
GHz in the X2.2 flare. The left panel of Fig.2 is the spectrogram
of the ZP structure observed at SBRS/Huairou. It shows that the
frequency range of the structure is occurred from 6.40 to 7.00 GHz
with a central frequency of 6.70 GHz. The time interval is from
01:49:50.2 to 01:49:51.5 UT, just at about 4 minutes after the
onset of the flare (01:46 UT), and about 6 minutes before the GOES
flare peak (01:56 UT). The duration of the ZP structure lasts for
about 1.3 s. It is composed with 3 stripes in distorted sinusoidal
wave shape arrayed on the longitudinal direction. The whole
sinusoidal wave shape drifts slowly to the low frequency, and the
drifting rate is about -300 MHz/s. The frequency bandwidth of the
zebra stripes is about 40 $\sim$ 60 MHz. The frequency separation
between the adjacent zebra stripes is about 80 $\sim$ 120 MHz, and
the relative frequency separation is $\Delta f/f\simeq$ 1.17 --
1.76 \%, increases slowly with respect to time. The emission of
the structure is strongly left-handed circular polarization with
polarization degree (defined as $pol=\frac{R-L}{R+L}\times 100\%$,
here R and L are the emission flux subtracted the background
emission) close to 100\%. Additionally, the ZP presents superfine
structures: each zebra stripe consists of millisecond spikes which
are practically vertical bright lines in the spectrogram with
duration at the limit of the time resolution (5 ms) and the
bandwidth of about 40 -- 60 MHz.

\begin{figure}
\begin{center}
 \includegraphics[width=8.3cm, height=8.5cm]{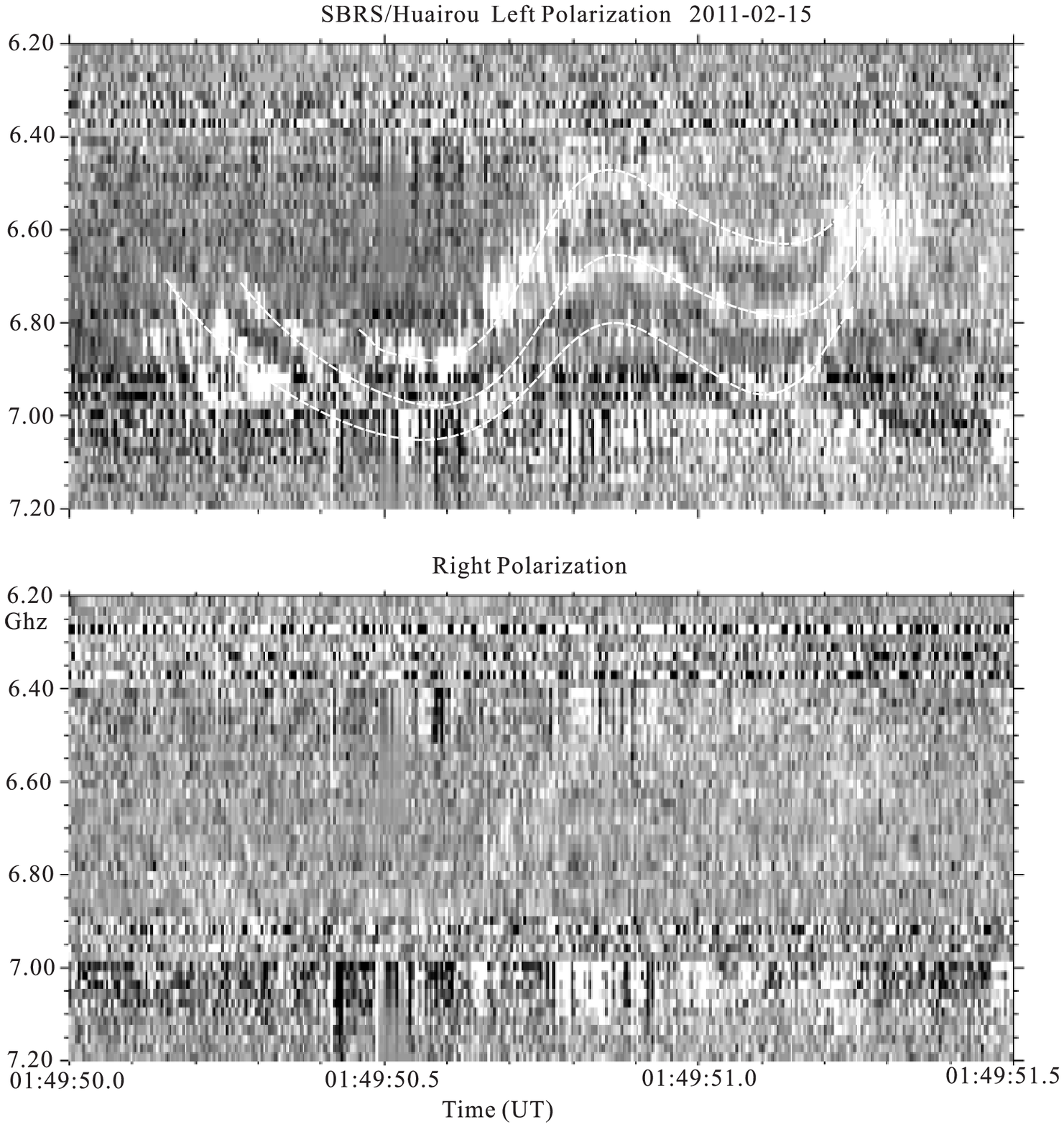}
  \includegraphics[width=7.6cm, height=8.2cm]{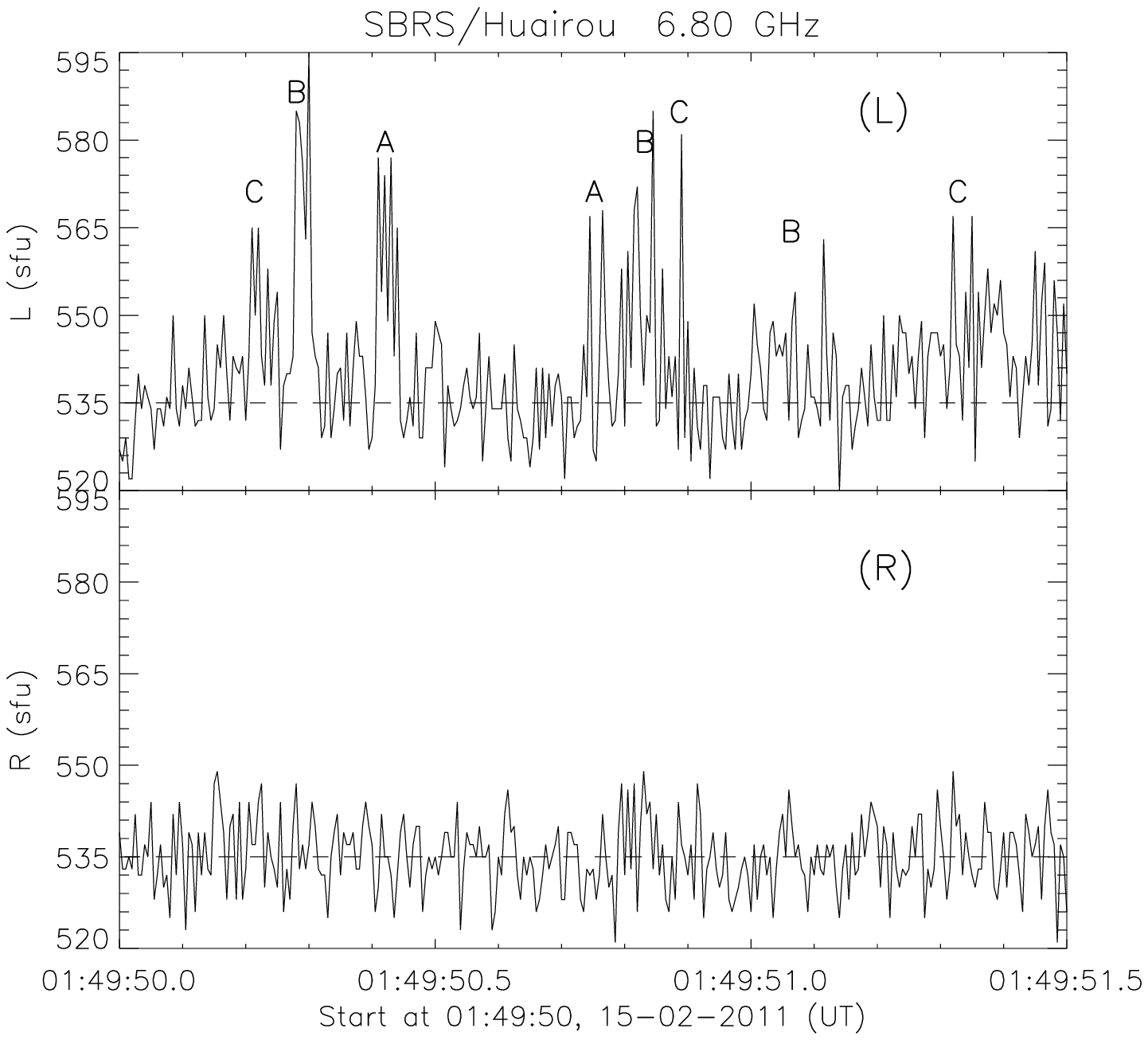}
  \caption{The left panel is the spectrogram of the Zebra pattern occurred at frequency of 6.40 $\sim$ 7.00 GHz observed at Chinese Solar
  Broadband Radiospectrometer (SBRS/Huairou) in 01:49:50 $\sim$ 01:49:51.5 UT, 15 Feb. 2011, the white dashed curves outshine the zebra stripes.
  The right panel is the profile of emission flux at frequency of 6.80 GHz in the same time interval of the Zebra pattern. A, B, and C
  represent respectively the stripes at high, middle, and low frequencies, which intersect with the horizontal level at frequency of 6.80 GHz in the left panel.}
\end{center}
\end{figure}

The right panel of Fig. 2 is the temporal profile of the microwave
flux at frequency of 6.80 GHz associated with the ZP1 structure.
In this figure, the averaged level of the ambient background
emission before and after the ZP structure is also presented (the
dashed line) which is about 535 sfu at left- and right-handed
circular polarization. The intensity of the zebra stripes have
enhancements of about 40 $\sim$ 80 sfu with respect to the ambient
emissions, which is much higher than the instrument sensitivity
(here $2\% S_{\bigodot}$ is about 4 $\sim$ 5 sfu at frequency of
6.20 $\sim$ 7.20 GHz). The positions of stripes are marked with
letters (A, B and C, which represent respectively the stripes at
high, middle, and low frequencies). At the same time, Fig. 2 is
also implying another feature: the emission intensity on the zebra
stripes decreases from the low frequency stripe to the high
frequency stripes.

\begin{figure}
\begin{center}
 \includegraphics[width=8 cm]{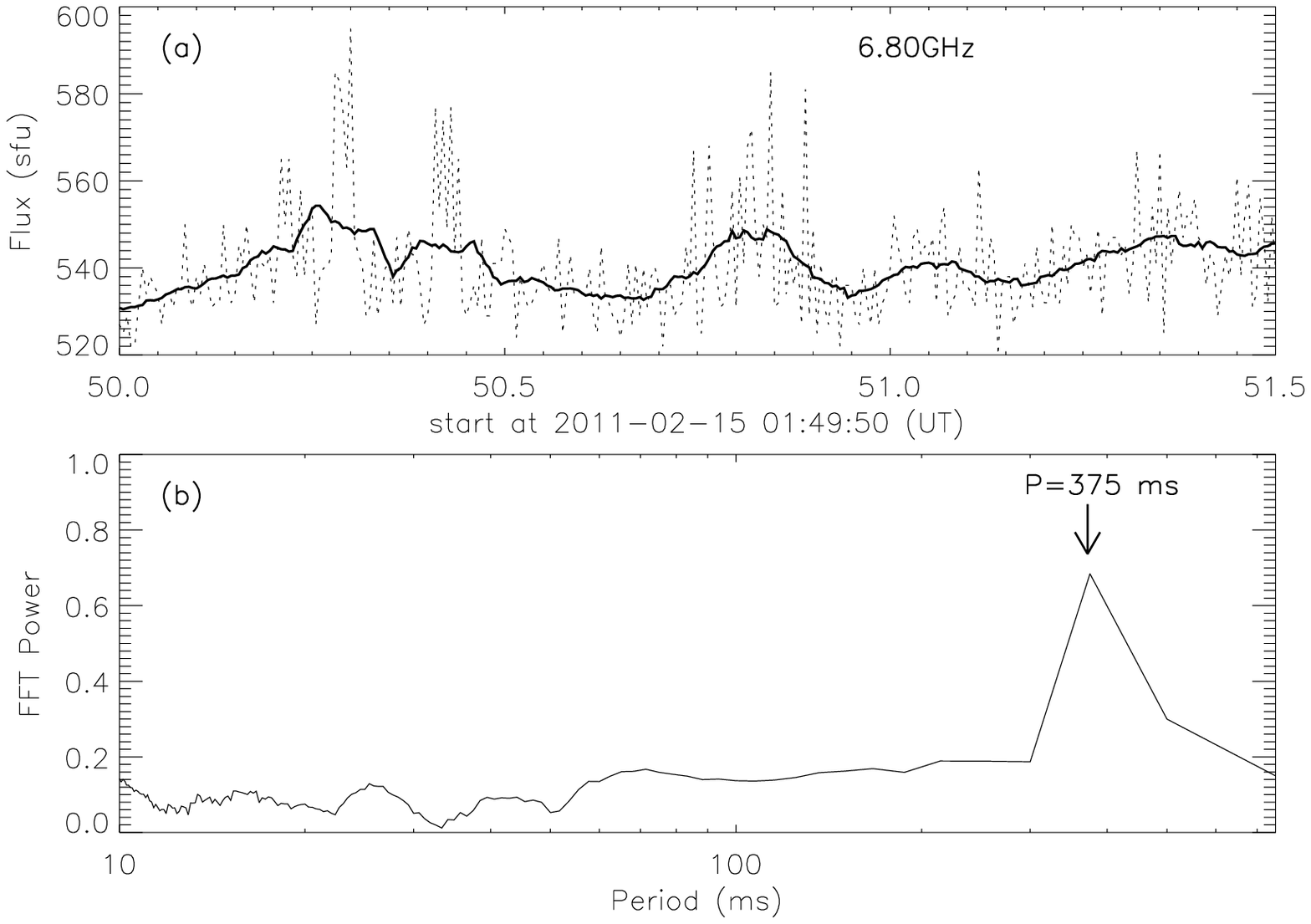}
  \caption{The FFT analysis on the microwave emission at frequency of 6.80 GHz, the upper panel is the profile of emission intensity,
  and the lower panel is the spectral power of the FFT analysis.}
\end{center}
\end{figure}

Fig.2 indicates that the zebra stripes of ZP1 are in distorted
sinusoidal wave shapes. If we suppose that the pattern is a
mixture of a general ZP structure and a quasi-periodic pulsation
(QPP), then it is easy to understand the distorted sinusoidal wave
shapes. Fig. 3 is the result of Fast Fourier Transformation (FFT)
analysis on the microwave emission at frequency of 6.80 GHz, which
indicates that the period of QPP is about 375 ms. This QPP belongs
to a very short-period pulsation (VSP).

(2) Zebra Pattern Structure at 2.60 $\sim$ 2.75 GHz

The second ZP structure is a weakly one (marked as ZP2 in Fig. 1)
at frequency of 2.60 $\sim$ 2.75 GHz in 02:01:19 $\sim$ 02:01:21
UT on the left-handed circular polarization spectrogram observed
at SBRS/Huairou (Fig. 4). There are 2 stripes in this structure
which can be discriminated from the spectrogram. The frequency
separation between the adjacent zebra stripes is 60 -- 70 MHz, and
the relative frequency separation is about $\Delta f/f_{0}\simeq$
2.23 -- 2.60 \%. The ZP structure lasts for about 1.5 s. It is
very weak, that the emission flux at the zebra stripes is only 15
$\sim$ 25 sfu higher than that of the background emissions. In the
first half part of the ZP structure, the frequency drift rate is
about 26 $\sim$ 42 MHz/s, with averaged value is about 35 MHz/s.
And in the second half part, the frequency drift rate reverses to
about -67 $\sim$ -90 MHz/s, with averaged value about -78 MHz/s.
The averaged value of the drift rate in the whole structure is
approximated to 0. From Fig. 4, it is reasonable to assume that
the ZP structure would be extended to the frequency range lower
than 2.60 GHz and the stripe number should be $>2$. However, as
lack of observation at the lower frequency range, we couldn't
confirm this conjecture.

\begin{figure}
\begin{center}
 \includegraphics[width=9.0 cm]{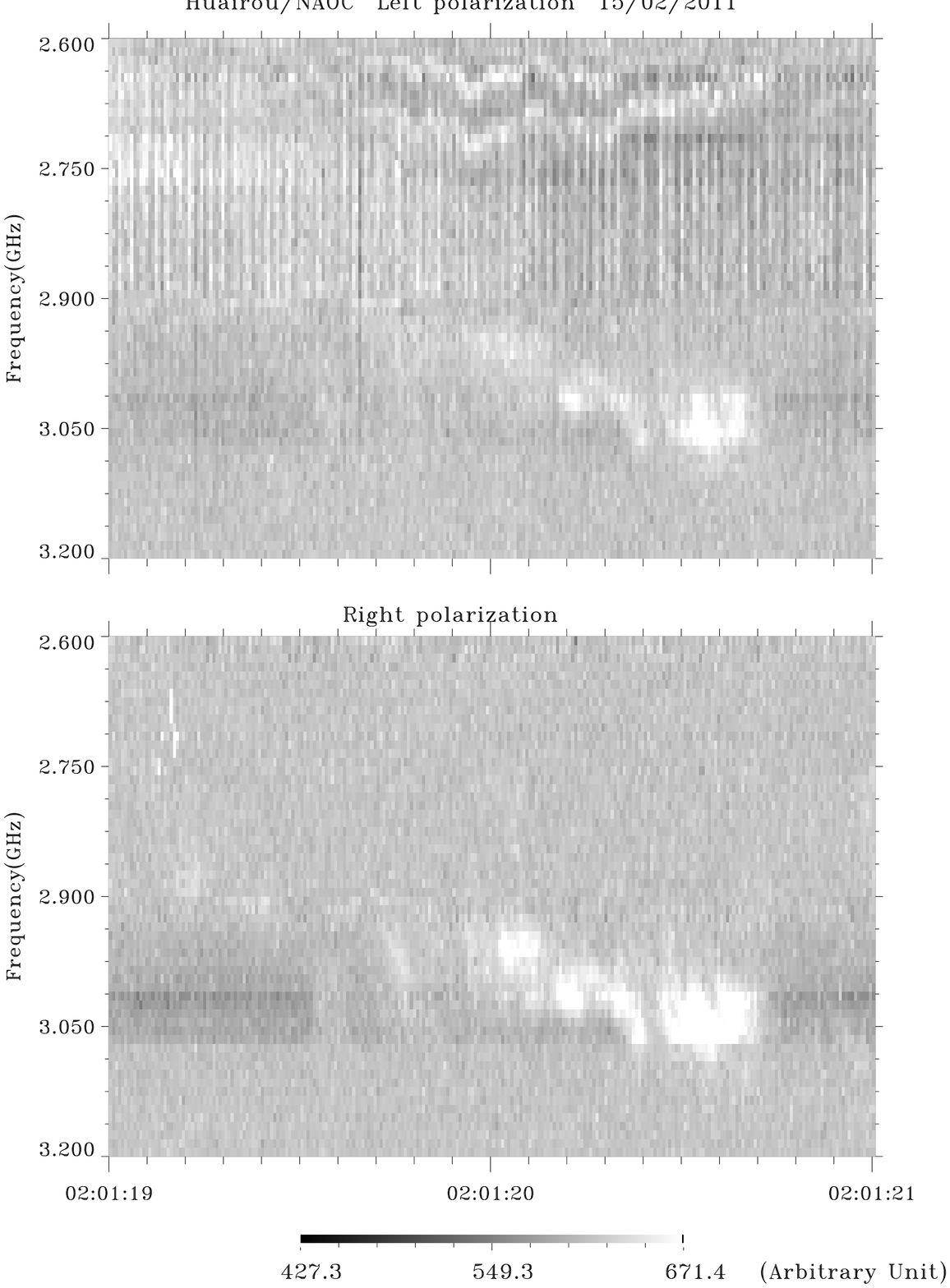}
  \caption{The spectrogram of a zebra pattern structure at frequency of 2.60 $\sim$ 2.75 GHz observed at Chinese Solar
  Broadband Radiospectrometer (SBRS/Huairou) in 02:01:19 $\sim$ 02:01:21 UT, on 15 Feb. 2011.}
\end{center}
\end{figure}

From the background of ZP2 on the spectrogram, a fast
quasi-periodic pulsation is also registered with frequency band of
2.62 -- 2.90 GHz, and the period of about 30 ms. On the high
frequency side of the ZP structure, there is a group of dot bursts
occurred at 2.92 -- 3.09 GHz, which frequency bandwidth is about
75 -- 85 MHz, and the polarization degree is $pol\sim$ 30 -- 35 \%
at right-handed circular polarization, and the lifetime of single
dot burst is in the range of 0.12 -- 0.28 s. The distribution of
the dots has a frequency drift rate of 180 MHz/s.

There are several clusters of fiber bursts on the spectrogram
occurred after the flare peak at frequency of 2.60 $\sim$ 3.00
GHz. One example of fiber burst occurred in 02:03:34 $\sim$
02:03:43 UT observed at SBRS/Huairou. The frequency drift rates of
the fibers are in the range of -211.9 MHz/s $\sim$ -403.2 MHz,
with averaged value of -303.5 MHz/s. This frequency drift rate is
much faster than the previous observations (e.g., the value was in
the range of 42.3 -- 87.4 MHz/s in the observations at the similar
frequency range of Wang \& Zhong, 2006, etc.). The central
frequency is about 2.80 GHz. The polarization of the fiber
emission is strong left-handed circular polarized, and the
polarization degree is $pol\sim$90\%.

Two type III bursts is also observed at SBRS/Huairou. The first
type III burst occurred at frequency of 2.62 $\sim$ 2.90 GHz in
02:07:50.70 $\sim$ 02:07:50.85 UT, the frequency bandwidth is 280
MHz, its frequency drift rate is in range of 11.25 $\sim$ 11.67
GHz/s. The emission is weakly right-handed circular polarization
with $pol\simeq$ 15 $\sim$ 20\%. The second type III burst
occurred at frequency of 2.64 $\sim$ 2.93 GHz in 02:07:51.55
$\sim$ 02:07:51.67 UT, the frequency bandwidth is 290 MHz, its
frequency drift rate is around 12.81 GHz/s, and it is also weakly
right-handed circular polarization with $pol\simeq$ 20-25\%.

(3) Zebra Pattern Structure at 1.04 $\sim$ 1.13 GHz

There is an obvious ZP structure occurred at 1.04 $\sim$ 1.13 GHz
in 02:10:56.8 $\sim$ 02:11:00 UT on 15 Feb. 2011, which is 15
minutes after the flare peak time, observed by SBRS/Yunnan, marked
as ZP3 in Fig.1. Fig.6 presents the spectrogram, which indicates
that there are 5 stripes in the ZP structure, and the duration is
about 3.2 s. In the first half part of the structure (before
02:10:58.6 UT), the frequency drift rate approximates to 0, but in
the second half part of the structure (after 02:10:58.6 UT) the
frequency drift rate is about -98 MHz/s. The frequency separation
between the adjacent zebra stripes is 14 $\sim$ 16 MHz, and the
relative frequency separation of the zebra stripes is about
$\Delta f/f\simeq$ 1.3 \%. The emission of the ZP structure is
mildly right-handed circular polarization, with polarization
degree $pol\sim$ 35 -- 40\%.

\begin{figure}
\begin{center}
 \includegraphics[width=8 cm]{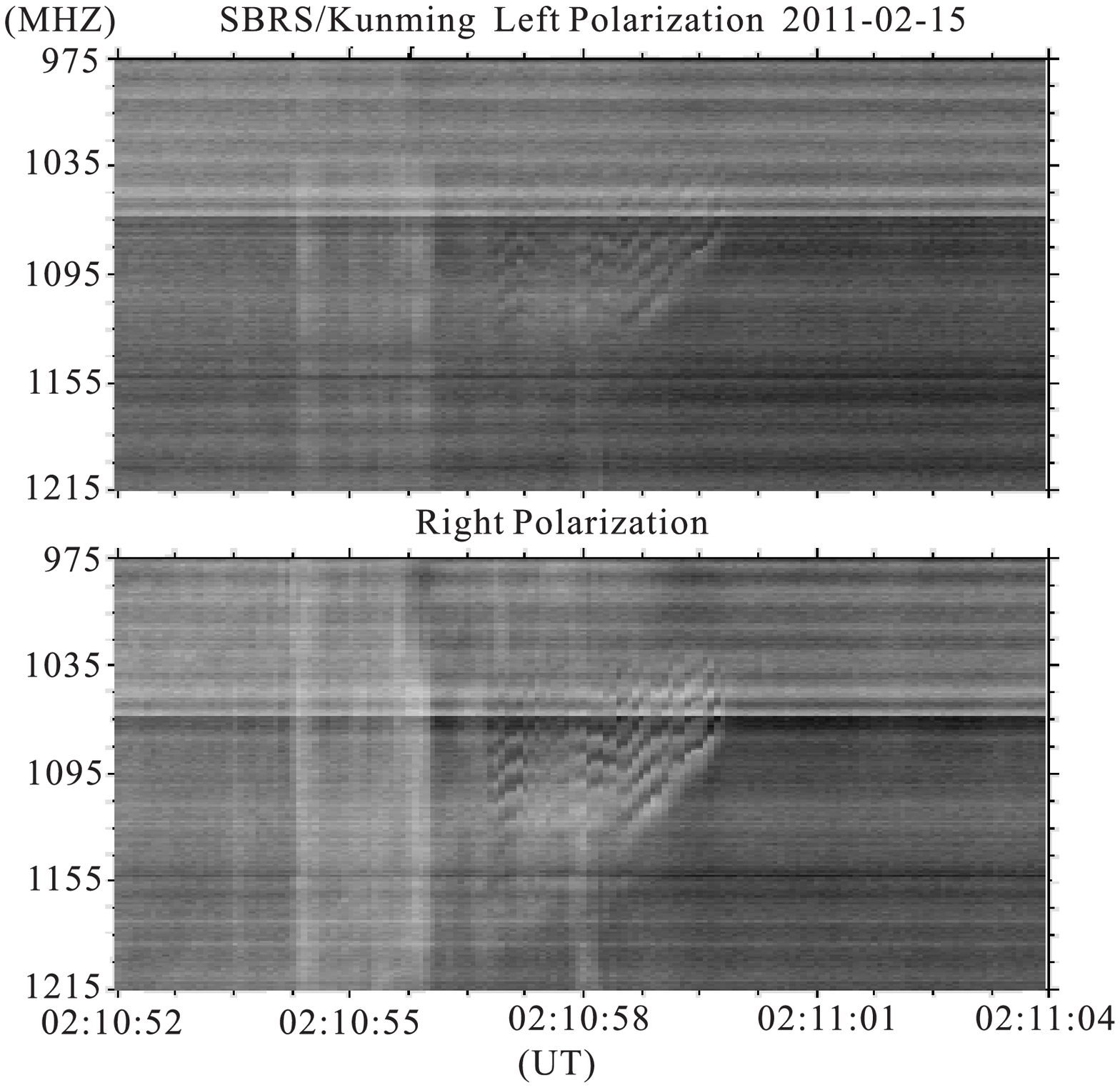}
  \caption{The spectrogram of a zebra pattern structure in the frequency of 1.04 $\sim$ 1.13 GHz observed by the Solar
  Broadband Radiospectrometer at Yunnan (SBRS/Yunnan) in 02:10:56.8 $\sim$ 02:11:00 UT, on 15 Feb. 2011.}
\end{center}
\end{figure}

Recently, a similar ZP structure was registered at frequency of
1.20 -- 1.40 GHz from the observations of the Frequency-Agile
Solar Radio-telescope Subsystem Testbed (FST) and the Owens Valley
Solar Array (OVSA) in the decay phase of an X1.5 flare on 14
December 2006 (Chen et al, 2011). The main properties (central
frequency, frequency separation, and the frequency drifting rate,
etc.) are very similar with ZP3 of this work.

Fig.6 presents the features of the image of extreme ultraviolet
171 \AA~ observed at SDO/AIA and the solar radio intensity
contours of the left (dashed) and right (solid) handed
polarizations at frequency of 17 GHz observed at NoRH just one and
a half minutes before ZP1, and only two minutes after the onset of
the flare. From this figure, we find that flare eruptive process
started from several separated small regions which behave as
several small discrete bright points on the image at extreme
ultraviolet 171 \AA~ observed at SDO/AIA. And the microwave
emission with maximum intensity at 17 GHz also distributed close
to the small discrete bright points. These facts indicate that the
magnetic reconnection and the energy release may break out from
several places with small size. ZP1 just times at this rising
phase, where most of the stored magnetic energy has not released,
and the magnetic field in the active region keeps in strong
status. The plasma in the magnetic loops may become very dense
because of the confinement of strong magnetic field. The magnetic
reconnection accelerates electrons to form anisotropic energetic
electron beams, this electron beam can excite low frequency
electrostatic waves, then couple with the upper-hybrid plasma
waves, and forms the ZP structures.

\begin{figure}
\begin{center}
 \includegraphics[width=8 cm]{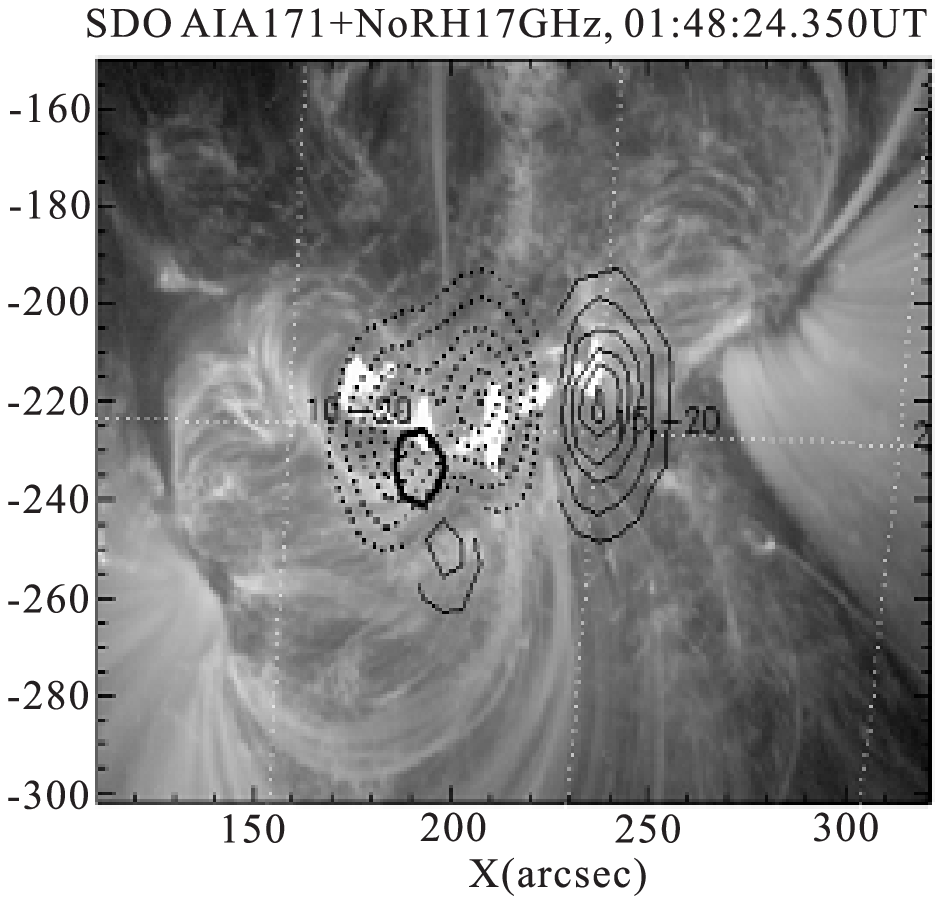}
  \caption{The solar microwave intensity contours of the left (dashed) and right (solid) handed polarizations at frequency of 17 GHz
  observed by NoRH overlapped on the background image of extreme ultraviolet 171 \AA~ observed by SDO/AIA.}
\end{center}
\end{figure}

Additionally, from Fig.1 we may find that the temperature
associated with the source region plasma is about 22 MK, 16 MK,
and 15 MK corresponding to the ZP structures occurred at frequency
of 6.40 -- 7.00 GHz, 2.60 -- 2.75 GHz, and 1.04 -- 1.13 GHz,
respectively.

Table 1 is the brief summary of the ZP structures observed in this
flare events. Here, the method of derived plasma density will be
introduced in next section. We find that the frequency separation
between zebra stripes increases with respect to the central
emission frequency, and the number of zebra stripes decreases with
respect to the central emission frequency.

\begin{table}\def~{\hphantom{0}}
  \begin{center}
  \caption{Main features of the ZP structures associated with the X2.2 flare. $f_{0}$ is the central frequency, DR is the frequency
  drift rate, pol is the polarization degree, $\Delta f$ is frequency separation of the zebra stripes, $\frac{\Delta f}{f_{0}}$ is the relative frequency separation
  of the zebra stripes, T is the temperature induced from the GOES soft X-ray emission.}
  \label{tab:kd}
  \begin{tabular}{lccccccccccccc}\hline
  ZP event                       &  ZP1                 & ZP2                  & ZP3                 \\\hline
  Start (UT)                     &  01:49:50.2          & 02:01:19.0           & 02:10:56.8          \\
  Duration (s)                   &  1.3                 &  1.5                 & 3.2                 \\
  Frequency Range (GHz)          &  6.40 -- 7.00        &  2.60 -- 2.75        & 1.04 -- 1.13        \\
  $f_{0}$ (GHz)                  &  6.70                &  2.68                & 1.08                \\
  DR (MHz/s)                     &  -300                &  +35 $\sim$ -78      & 0 $\sim$ -98        \\
  pol (\%)                       &  90 -- 100           &  90 -- 100           & 35-- 40             \\
  $\Delta f$ (MHz)               &  80$\sim$120         &  60$\sim$70          & 14$\sim$16          \\
  $\frac{\Delta f}{f_{0}}$(\%)   &  1.17 -- 1.76        &  2.23 -- 2.60        & 1.30 -- 1.48        \\
  stripes                        &  3                   &  $>$2                & 6                   \\
  plasma density~($cm^{-3}$)     &  $5.5\times10^{11}$  &  $8.8\times10^{10}$  & $1.4\times10^{10}$  \\
  T (MK)                         &  22                  &  16                  & 15                  \\\hline
  \end{tabular}
 \end{center}
\end{table}

In an ordinary ZP structure, the frequency separation between the
adjacent stripes grows with the emission frequency: from 4 $\sim$
5 MHz at 200 MHz to about 80 MHz at 3.0 GHz, and to about 150
$\sim$ 200 MHz at 5.70 GHz, and the relative frequency separation
of stripes is about a constant: $\Delta f/f\simeq$ 2 $\sim$ 3.5
\%. However, in this X2.2 flare event, we find that ZP1 and ZP3
has a narrow relative frequency separation of the zebra stripes,
which is smaller than 1.8\%, although the relative frequency
separation of ZP2 stripes is very close to the above general ZP
structure.

\begin{table}\def~{\hphantom{0}}
  \begin{center}
  \caption{Lists of the frequency, frequency ratio of the adjacent zebra stripes in each ZP structure.}
  \label{tab:kd}
  \begin{tabular}{lccccccccccccc}\hline
  ZP event    &  Stripe No. & Frequency(GHz) & Frequency Ratio    \\\hline
  ZP1         &  1             & 6.80           &                 \\
              &  2             & 6.66           & 1.021           \\
              &  3             & 6.54           & 1.018           \\\hline
  ZP2         &  1             & 2.73           &                 \\
              &  2             & 2.65           & 1.030           \\\hline
  ZP3         &  1             & 1.124          &                 \\
              &  2             & 1.107          & 1.015           \\
              &  3             & 1.094          & 1.012           \\
              &  4             & 1.080          & 1.013           \\
              &  5             & 1.066          & 1.013           \\\hline
  \end{tabular}
 \end{center}
\end{table}

Table 2 presents the frequency ratio of the adjacent zebra stripes
in each ZP structure. We find that the frequency ratios of the
adjacent zebra stripes are approximated to a constant of about
1.020 in ZP1, and 1.013 in ZP3. This feature is much more obvious
in ZP3 where there are 5 zebra stripes, the maximum ratio is 1.015
while the minimum ratio is 1.012, all of them are very close to
1.013. This fact indicates that the Ledenev's model is not
suitable to interpret the formation of the ZP structures observed
in this work.

\section{Physical Discussion on the Zebra Patterns}

At first, almost all the theoretical models (BM model, WW model,
or DPR model) for the generation of ZP structures indicate that
the ZP emission frequencies are approximately around the plasma
frequency or its second harmonics. With this point, the plasma
density in the ZP source regions can be estimated: $n_{e}\approx
f^{2}/81s^{2}$. As ZP1 and ZP2 have strongly polarization, $s=1$.
As for ZP3, its has moderate polarization degree (35 -- 40\%).
According to the calculation of Dulk (1985), the polarization
degree of the second harmonic plasma emission is only 10 -- 20\%.
So, ZP3 also possibly belongs to the fundamental emission $s=1$.
Substituting the emission frequencies of ZP structures into this
expression, the range of plasma density in the ZP source regions
can be obtained, respectively:

ZP1, $5.1\times10^{11}$ -- $6.0\times10^{11}$ cm$^{-3}$, the
averaged value is $5.5\times10^{11}$ cm$^{-3}$;

ZP2, $8.3\times10^{10}$ -- $9.3\times10^{10}$ cm$^{-3}$, the
averaged value is $8.8\times10^{10}$ cm$^{-3}$;

ZP3, $3.3\times10^{9}$ - $4.0\times10^{9}$ cm$^{-3}$, the averaged
value is $1.4\times10^{10}$ cm$^{-3}$.

Generally, the source region with plasma density as up to
$5.5\times10^{11}$ cm$^{-3}$ is always located very close to the
base of solar corona where the height from the solar photosphere
is only several thousands kilometers, while the source region with
plasma density of about $\times10^{10}$ cm$^{-3}$ is located near
the bottom of the solar corona. However, it should be different
around the active regions, especially around the flaring regions.
The X-ray observations indicate that the plasma densities around
the flaring core region are in the range of $10^{9} - 10^{11}$
cm$^{-3}$, and their heights can be in several decades of
thousands km above the solar photosphere (Ohyama \& Shibata,
1998).

One of the crucial and most difficult problem in solar physics is
to determine the coronal magnetic field confidently. There are
many publications which present the estimations of the coronal
magnetic field by using solar radio observations (Mann, Karlicky,
\& Motschmann, 1987; Gelfreikh, 1998; Huang \& Nakajima, 2002;
Huang 2008, etc.). Recent observations of microwave bursts with
fine structures open up a new possibilities for determining the
coronal magnetic field (Karlicky \& Jiricka, 1995; Lenedev et al,
2001, etc). The ZP structure is one of most important microwave
fine structures which can be used to diagnose magnetic field
strength in the coronal source regions, although the results
depend on the theoretical models. Different ZP model will deduce
different values of magnetic field in the ZP source region.
Practically, it is always difficult to verdict which model is the
best one fitted to observations. Possibly, from the estimations of
the magnetic field strengths from the ZP structures, we could get
a considerable restriction for the theoretical models.

(1) BM model indicates that the frequency separation of the
adjacent zebra stripes is just equal to the electron
gyro-frequency. From this we may obtain a direct measurement of
the magnetic field in the coronal source region:

\begin{equation}
B\simeq \frac{2\pi m_{e}}{e}\Delta f\simeq 35.6\times10^{-8}\Delta
f.
\end{equation}

Here, the unit of $B$ is in Gauss, and $f$ in Hz. Substituting the
frequency separation of ZP1, ZP2, and ZP3 into the above
expression, we may get the magnetic field strength as 28 -- 43 G,
21 -- 25 G, and 5 -- 6 G, respectively. In this regime, the
magnetic field strength is only depending on the frequency
separation between the adjacent zebra stripes.

(2) From WW model, we may get the magnetic field strength in ZP
source region:

\begin{equation}
B\simeq 2\frac{2\pi m_{e}}{e}\Delta f\simeq
71.2\times10^{-8}\Delta f
\end{equation}

With this relation, the magnetic field strength is two times of
that estimated from BM model: 55 -- 85 G, 42 -- 49 G, and 10 -- 11
G, corresponding to ZP1, ZP2, and ZP3, respectively. This regime
is also independent to the inhomogeneous scale height in the
source region.

(3) From DPR model, we may obtain the measurement of magnetic
field strength in the ZP structure source region. Based on
Equation (3) and (4), the magnetic field strength can be derived:

\begin{equation}
B\simeq \frac{2\pi m_{e}}{e}\cdot Q\cdot\Delta f.
\end{equation}

Here, $Q$ is an inhomogeneous factor which is dominated mainly by
the scale heights of plasma density $n_{e}$ and the magnetic field
$B$ in the source region. It can be expressed as:

\begin{equation}
Q=\frac{|2H_{n}-H_{b}|}{sH_{b}}.
\end{equation}

Here $s$ is the harmonic number. When the emission generated from
the coalescence of an excited plasma wave and a low frequency
electrostatic wave, the emission frequency equals nearly to plasma
frequency, $s=1$ (fundamental emission); and when the emission
generated from the coalescence of two excited plasma waves, the
emission frequency equals nearly to the double plasma frequency,
$s=2$ (second harmonics).

The scale heights of plasma density $H_{n}$ and the magnetic field
$H_{b}$ are two crucial parameters which control the magnitude of
$Q$. However, they are depending on the atmospheric models around
the source region. We may adopt the Newkirk model to express the
plasma density around the source region:
$n(r)=M\times10^{\frac{4.32}{r}}$. $M$ is a constant in the
Newkirk model which may change depending on the different coronal
regions. However, it does't change the following estimations when
we change the magnitude of $M$ (it doesn't appear in the
expression of $H_{n}$). The scale height of plasma density can be
deduced:

\begin{equation}
H_{n}\simeq 70r^{2}, (Mm).
\end{equation}

As for the coronal magnetic field above active region, the Dulk \&
McLean model is always adopted: $B(r)=0.5m/(r-1)^{1.5}, (1.02 \leq
r\leq 10)$ (Dulk \& McLean, 1978). However, this model represents
only an averaged decrease of the magnetic field strength with
height, the real values may deviate by a factor $m$ of up to 3.
$r$ is the height from the solar center in unit of the solar
optical radius $R_{\odot}$.

More precisely, we may take the coronal loop model to obtain the
scale height of the magnetic field in ZP source regions, the
magnetic field strength in a coronal loop can be written as:
$B(h)\sim B_{0}(1+h/d)^{-3}$ (Takakura, 1972). Here, $B_{0}$ is
the magnetic field at the foot-point of the loop, $d$ is the
distance between the two foot-points, and $h$ is the height from
the photosphere. Then the scale height of magnetic field can be
deduced as:

\begin{equation}
H_{b}=\frac{d}{3}(1+\frac{h}{d}).
\end{equation}

According to the work of Maurya, Reddy, \& Ambastha (2011), we
know that the maximum magnetic field on the sunspot is about 1400
G. From this figure, we also find that the distance between the
maximum centers of left- and right-handed polarizations at
frequency of 17 GHz is only about 40 arc-seconds, approximated to
29 Mm. Because the maximum centers of left- and right-handed
polarizations at frequency of 17 GHz are close to the foot-points
of the plasma loop, we may assume $d\sim$ 29 Mm.

Equ (8) -- (11) indicate that the magnetic field strength depends
on another unbeknown key parameter: $r$ or $h$, here,
$h=(r-1)R_{\odot}$. As lack of the imaging observation at the
corresponding frequencies, we may give some assumptions.
Considering the frequency of ZP1 (6.40 -- 7.00 GHz) and the
previous works (review of Gary \& Keller, 2004), its source region
is very close to the core region of the flaring energy release, we
may assume its height is about 20 Mm (above the solar
photosphere). Then the heights of ZP2 and ZP3 can be deduced as 39
Mm and 61 Mm by Newkirk model and plasma emission mechanism. With
above estimations, assumptions and observations, we may obtain the
magnetic fields in the source regions of ZP1, ZP2 and ZP3 are 230
-- 345 G, 126 -- 147 G and 23 -- 26 G, respectively. The values of
ZP1 and ZP2 are very close to that estimated from Takakura model,
and also close to that estimated from Dulk \& McLean model when
the factor $m=3$. However, the value of ZP3 is only approaching to
that estimated from Dulk \& McLean model when the factor $m=1$,
much smaller than that estimated from Takakura model.

Additionally, there are several clusters of fiber bursts at
frequency of 2.60 -- 3.00 GHz which is very close to the frequency
band of ZP2. The fiber bursts have been thought to be another
promising way to diagnose the coronal magnetic field in the source
region. It can be generated by packets of low frequency whistler
waves propagating along the magnetic field lines of coronal loop.
From the frequency drift rate, we can derive the magnetic field
(Benz \& Mann, 1998):

\begin{equation}
B\simeq \frac{\pi H_{n}m_{e}}{ec\sqrt{x}\cdot
cos\theta}\frac{df}{dt}\approx
5.93\times10^{-20}\frac{H_{n}}{\sqrt{x}cos\theta}\frac{df}{dt},
(Tesla)
\end{equation}

$H_{n}$ can be calculated from Equ.(10) in about 78 Mm.
$x=\frac{f_{w}}{f_{ce}}$ is the ratio between whistler frequency
and the electron gyro-frequency, $\theta$ is the declining with
respect to the magnetic field line. Generally, $\theta\sim 0$, and
$x\sim 0.01$, then we may substitute the frequency drift rates of
the fiber burst at frequency of 2.60 -- 3.00 GHz, the magnetic
field strength can be obtained as 98.6 -- 187.7 G, with the
averaged value of 143.1 G. This value covers the result estimated
by DPR model of ZP2. It can be regarded as a collateral evidence
for the estimations of magnetic field in the ZP source regions
from DPR model.

\begin{table}\def~{\hphantom{0}}
  \begin{center}
  \caption{Magnetic field strengths (unit in Gauss) estimated from different models in ZP source regions.}
  \label{tab:kd}
  \begin{tabular}{lccccccccccccc}\hline
  ZP event                       &  ZP1       &  ZP2       &    ZP3    \\\hline
  ZP BM model                    &  28 -- 43  &  21 -- 25  &  5 -- 6   \\
  ZP WW model                    &  55 -- 85  &  42 -- 49  & 10 -- 11  \\
  ZP DPR model                   & 230 -- 345 & 126 -- 147 & 23 -- 26  \\\hline
  Dulk \& McLean model: m=1      &    102     &     38     &    19     \\
  ~~~~~~~~~~~~~~~~~~~~~ m=2      &    204     &     76     &    38     \\
  ~~~~~~~~~~~~~~~~~~~~~ m=3      &    306     &     114    &    57     \\
  Takakura model                 &    290     &    109     &    48     \\\hline
  \end{tabular}
 \end{center}
\end{table}

Table 3 presents the magnetic field strengths obtained from
different models in ZP source regions. As a comparison, we also
listed the estimations from Dulk \& McLean model ($m=$ 1, 2, and
3) and Takakura model at the corresponding heights, respectively.
We find that BM model and WW model present very low estimations of
the magnetic field in ZP source regions; Dulk \& McLean model
gives a low estimations for ZP1 and ZP2 when m=1 or 2. The result
of DPR model gets the furthest consistency with the Takakura model
and the Dulk \& McLean model with m=3 when estimate the magnetic
fields in source regions of ZP1 and ZP2. Especially in ZP2,
estimations of the magnetic field from DPR model and Takakura
model are also consistent with the result obtained from fiber
bursts at the similar frequency range. So, we prefer to suppose
that the DPR model is possibly the real model for the ZP
structures observed in this work. However, there is a 2 times
difference between DPR model and Takakura model in ZP3 source
region. It may just reflect that the Takakura model is valid only
within a magnetic loop. In our above estimations, the height of
the ZP3 source region is about 61 Mm, which is more than two times
of distance ($d\sim$ 29 Mm) between the two foot-points of the
loop, such height should be much higher than the loop-top. It is
possible that the source region of ZP3 is located at another
different magnetic loop with a larger $d$ and a much low $B_{0}$.
As lack of such observations, we can not confirm such inference.

In section 2 we have pointed out that ZP1 is possibly a mixture of
a general ZP structure and a QPP, and the QPP is a very
short-period pulsation (VSP). From the work of Tan, et al (2007)
and Tan, et al (2010), we may suppose that the QPP is possibly a
result of modulations of the resistive tearing-mode oscillations
in the current-carrying flare plasma loops with high temperature.
The panel (9) of Fig.1 indicates that temperature around ZP1 22
MK) is very close to the maximum of the profile, it is possible
that the modulations of the resistive tearing-mode oscillations
take place. The current-carrying plasma loop can drive the
tearing-mode oscillation and modulate the microwave emission to
form VSP. On this VSP background, some mechanism generate the ZP
structure.

\section{Conclusions}

On 15 February 2011, there erupts an X2.2 flare event on the solar
disk, which was the first X-class flare occurred since the solar
cycle 24. Associated with this flare event, three microwave ZP
structures at different frequencies are registered in different
phases of the flare: the first is registered from SBRS/Huairou at
frequency of 6.40 $\sim$ 7.00 GHz, which is very unusual at such
high frequency band and in the early rising phase of the flare;
the second is also registered from SBRS/Huairou, at frequency of
2.60 -- 2.73 GHz in the decay phase of the flare, possibly it may
extend to the frequency lower than 2.60 GHz; the third is
registered from SBRS/Yunnan at frequency of 1.04 -- 1.13 GHz, in
the decay phase after far from the flare peak.

By scrutinizing the current prevalent theoretical models of ZP
structures (including Bernstein model, whistler wave model, DPR
model, and the Ledenev model), comparing their estimated magnetic
field strengths in the corresponding source regions, we find that
the DPR model is much more possible for explaining the generation
of microwave ZP structures. It derived the magnetic field
strengths as about 230 -- 345 G, 126 -- 147 G, and 23 -- 26 G in
the source regions of ZP1, ZP2, and ZP3, respectively. Comparison
with the diagnostics of fiber bursts and the previous empirical
model, we suggest that such estimations are acceptable.

It should be noted that DPR model is not self-contained when we
adopt it to diagnose the magnetic fields in ZP source regions. It
needs the supplement of the inhomogeneity model of plasma density
and magnetic field. However, it is not easy to get the exact
inhomogeneity models. So far, all the existing models (e.g. Dulk
\& McLean model, Takakura model, etc ) of plasma density and
magnetic field are proposed that the plasma density and magnetic
field change with the height, and expressed as functions of height
($r$ or $h$). Such method implies that the inhomogeneous scale
heights of magnetic field is also a function of magnetic field,
and this is the origin of the self-contradictions. We need a more
perfect model which can provide the inhomogeneous scale lengths of
plasma density and magnetic field, and they are independent of the
magnitude of magnetic field strength. The magnetic field
diagnostics of BM model and WW model are independent of the
inhomogeneity models, they seem to be the perfect models to
diagnose the magnetic field in the source region by ZP structures.
But they are not likely to agree with the ZP structures observed
in this work.

However, we should be noted that either the Dulk \& McLean model,
or the Takakura's model is only a simplified model. The actual
regime during the microwave burst with ZP structures should be
extremely dynamic processes, the real magnetic topology is also
much more complex and changeable than the depiction of the above
models. So far, the only thing we can do is to obtain the
approximated estimations. The relative large difference of the
magnetic estimations in ZP3 source region between DPR model and
the Takakura's model just reflect that its source region should be
located at different magnetic loop with different distance of the
foot-points and different different initial magnetic field.

From the above discussions, we know that the exact inhomogeneity
models of magnetic field and plasma density are most important.
However, so far, because of lack of imaging observation with
spatial resolutions in the corresponding frequencies, it is
difficult to obtain the configuration features of the coronal
magnetic field in the source region. To overcome such problems, we
need some new telescopes, for example, the constructing Chinese
Spectral Radioheliograph (CSRH, 0.4 - 15 GHz, will be finished
before 2014) in the decimetric to centimeter-wave range (Yan et
al, 2009) and the proposed American Frequency Agile Solar
Radiotelescope (FASR, 50 MHz - 20 GHz) (Bastian, 2003). When these
instruments begin to work, we may obtain the solar radio
observations with high spatial-temporal resolutions at
multi-frequency channels. One of the most important way to measure
the coronal magnetic field is to probe ZP structures in each
subareas with high spatial resolutions, and deduce the magnetic
field from certain theoretical models. With these development, we
will get more perfect understanding of the elementary processes in
solar flares.

\acknowledgments

The authors would like to thank the anonymous referee for the
helpful and valuable comments on this paper. We would also thank
the the GOES, NoRP, NoRH, SDO/AIA, and SBRS/Huairou teams for
providing observation data. This work is mainly supported by NSFC
Grant No. 10733020, 10873021, 10921303, MOST Grant No.
2011CB811401, the National Major Scientific Equipment R\&D Project
ZDYZ2009-3, RFBR 10-02-00153 and 11-02-10000-k. This research was
also supported by a Marie Curie International Research Staff
Exchange Scheme Fellowship within the 7th European Community
Framework Programme. The research carried out by Sych Robert at
National Astronomical Observatories (NAOC) was supported by the
Chinese Academy of Sciences Visiting Professorship for Senior
International Scientists, grant No. 2010T2J24. Guannan Gao's work
is supported by CAS-NSFC Key Project(Grant No.10978006).

\end{document}